\documentclass[11pt]{article}
\usepackage{longtable}
\usepackage{times, subfigure}
\usepackage{alltt}
\usepackage{framed}
\usepackage{color}
\usepackage{harvard}
\usepackage{url}
\usepackage{epsfig,setspace,rotating,fullpage}
  
\title{
Using a ``Study of Studies" to help statistics students assess research findings}
\author{Azka Javaid \and Xiaofei (Susan) Wang \and Nicholas J. Horton\thanks{
Address for correspondence: Department of Mathematics and Statistics, Amherst College, AC\#2239, PO Box 5000,
Amherst, MA  01002-5000.  Phone: 413-542-5655, email:
nhorton@amherst.edu}  
}
\begin{document}
\maketitle
\newpage
\begin{center}
{\large
Using a ``Study of Studies" to help statistics students assess research findings}
\end{center}


\section*{Introduction}

The American Statistical Association's Undergraduate Guidelines Workgroup in the Curriculum Guidelines for Undergraduate Programs in Statistical Science emphasize the importance of study design as an essential skill needed for undergraduate programs in statistics \cite{asa:2014}. Communication skills are also stressed along with teamwork and collaboration as essential elements for statistical practice. In addition, the Guidelines for Assessment and Instruction in Statistics Education (GAISE) College Report notes the importance of concepts like bias and causal inference in study design \cite{gaise:2016}. To achieve these goals, the GAISE College Report includes several recommendations, one of which reiterates the importance of fostering active learning through discussions (p. 18-20).

This article describes an activity that is appropriate for students in introductory and intermediate statistics courses to practice interpreting research results and scrutinizing the design and analysis of studies. The activity includes a component of group work to improve students' communication and collaboration skills. 

Traditional textbooks used in the introductory statistics curriculum stress the importance of survey design. As an example, the fourth edition of \emph{Intro Stats} (IS) by Richard De Veaux, Paul F. Velleman and David E. Bock, reinforces concepts relevant to experimental design including bias, randomization and sample size in Chapter 12 (``Sample Surveys") \cite{IS4}. Additional sample design concepts like observational studies, control groups, statistical significance and confounding variables are presented in Chapter 13 (``Experiments and Observational Studies") of \emph{Intro Stats}. Other textbooks (e.g., \emph{OpenIntro Statistics}, \cite{openintro}) follow a similar approach. 

How can textbook readings be reinforced in a class? We describe an activity to help students explore aspects of design, assess research findings in published papers, and critique representations and interpretations of original research. 

\section*{The Activity}

\subsection*{Study of Studies}

``Study of Studies" is a column regularly published by the \emph{The Atlantic} magazine. Each ``Study of Studies" analyzes a different topic using published research articles, with full citations provided at the end. Table \ref{tab:diner} lists the name, author, and date for the all the past published ``Studies."

\begin{table}[htpb]
\begin{tabular}{|p{7.5cm}|p{3.5cm}|p{3.75cm}|}
\hline
Title & Author & Publication Date\\ \hline
You Can Be Too Beautiful & James Hamblin & March 2013 \\
The Queen Bee's Guide to Parenting & Lindsey Abrams & April 2013 \\
Various Ways You Might Accidentally Get Drunk & James Hamblin & May 2013 \\
The Unexpected Ways a Fetus Is Shaped by a 
Mother's Environment	& Lindsey Abrams & June 2013 \\
The Worst Time to Have Surgery & James Hamblin & July/August 2013 \\
Is There Really Such a Thing as a `Workaholic'? & Jordan Weissmann & September 2013 \\
Violence Is Contagious & Rebecca J. Rosen & October 2013 \\
Why You Look Like Your Dog & Sarah Yager & November 2013 \\
How Women Change Men & Sarah Yager & December 2013 \\
Who Cheats-and Why & Julie Beck & January/February 2014\\
Why You Can't Keep a Secret & Sarah Yager & March 2014 \\
The Optimal Office & Julie Beck & April 2014 \\
Our Gullible Brains & Sarah Yager & May 2014 \\
Funny or Die & Julie Beck & June 2014 \\
What is Art? & Matthew Hutson & July/August 2014 \\
How to Look Smart & Julie Beck & September 2014 \\
Status Anxiety & Matthew Hutson & October 2014 \\
Keeping the Faith & Emma Green & November 2014 \\
Faking It & Julie Beck & December 2014 \\
You Are Just Like Me! & Matthew Hutson & January/February 2015 \\
The Secret of Superstition & Matthew Hutson & March 2015 \\
Diner Beware & Bourree Lam & April 2015 \\
When Emotional Intelligence Goes Wrong & Andrew Giambrone	& May 2015 \\
The Hypocrisy of Professional Ethicists & Emma Green & June 2015 \\
Palm Reading Is Real? & Eleanor Smith & July/August 2015 \\
A Scientific Look at Bad Science & Bourree Lam & September 2015 \\
Why We Compete & Matthew Hutson & October 2015 \\
The Strange Origins of Urban Legends & Matthew Hutson & November 2015 \\
Why You Bought That Ugly Sweater & Eleanor Smith & December 2015 \\
A Strategic Guide to Swearing & Stephanie Hayes & January/February 2016 \\
People Are Pretty Bad At Reading Faces & Naomi Sharp & March 2016 \\
CEOs Behaving Badly & Alyza Sebenius & April 2016 \\
How to Boast on the Sly & Matthew Hutson & May 2016 \\
Life Isn't Fair & Matthew Hutson & June 2016 \\
The Science of Beer Goggles & Stephanie Hayes & July/August 2016 \\
The Charisma Effect & Matthew Hutson & September 2016 \\
Do People Need Small Talk to Be Happy? & Stephanie Hayes & October 2016 \\
How Voters Respond to Electoral Defeat & Ben Rowen & November 2016 \\
Why Kids Need Recess & Alia Wong & December 2016 \\
\hline
\end{tabular}
\caption{List of past ``Study of Studies" published by \emph{The Atlantic}}
\label{tab:diner}
\end{table}
    For this activity, we utilized one of the ``Study of Studies" articles titled ``Diner Beware: How restaurants trick you into eating less and spending more"
(\url{http://www.theatlantic.com/magazine/archive/2015/04/diner-beware/386267}) \cite{lam3:2015}. Author Bourree Lam analyzes how restaurants manipulate seating arrangement, server posture, plate color and size, and music to attract more customers and revenue. Twelve research articles are summarized in short excerpts with full citations provided as footnotes. Table \ref{tab:diner2} displays the list of papers included in the ``Diner Beware" column.

\begin{table}[htpb]
\begin{tabular}{|p{10cm}|p{5.2cm}|}
\hline
Title and Author & Publication \\ \hline
Odors and Consumer Behavior in a Restaurant [Gu\'eguen and Petr 2006] & \textit{International Journal of Hospitality Management} \\
Plate Size and Color Suggestibility [Ittersum and Wansink 2012] & \textit{Journal of Consumer Research} \\ 
Assessing the Influence of the Color of the Plate on the Perception of a Complex Food in a Restaurant Setting [Fiszman, Giboreau and Spence 2013] & \textit{Flavour} \\
Dining in the Dark [Scheibehenne, Todd and Wansink 2010] & \textit{Appetite} \\
The Effect of Musical Style on Restaurant Customers' Spending [North, Shilcock and Hargreaves 2003] & \textit{Environment and Behavior} \\
The Influence of Background Music on the Behavior of Restaurant Patrons [Milliman 1986] & \textit{Journal of Consumer Research}\\
The Impact of Restaurant Table Characteristics on Meal Duration and Spending [Kimes and Robson 2004] & \textit{Cornell Hotel and Restaurant Administration Quarterly} \\
How a Crowded Restaurant Affects Consumers' Attribution Behavior [Tse, Sin and Yim 2002] & \textit{International Journal of Hospitality Management} \\
Lower Buffet Prices Lead to Less Taste Satisfaction [Just, Sigirci and Wansink 2014] & \textit{Journal of Sensory Studies} \\
Determinants and Consequences of Female Attractiveness and Sexiness [Lynn 2009] & \textit{Archives of Sexual Behavior} \\
Effect of Server Posture on Restaurant Tipping [Lynn and Mynier 1993] & \textit{Journal of Applied Social Psychology} \\
Effect on Restaurant Tipping of Male and Female Servers Drawing a Happy, Smiling Face on the Backs of Customers' Checks [Rind and Bordia 1996] & \textit{Journal of Applied Social Psychology} \\
\hline
\end{tabular}
\caption{12 papers included in Bourree Lam's ``Study of Studies" on Restaurants and Dining (April, 2015 \textit{The Atlantic} Magazine)}
\label{tab:diner2}
\end{table}

\subsection*{Implementation}

The students were provided with copies of the one-page ``Diner Beware" column, which was read aloud by the class.  Next, they were split into groups of two to four students and each was provided a copy of one of the twelve research articles cited in ``Diner Beware." The research articles ranged in length from 4 to 28 pages, with an average of 10 pages. 

The students were asked to skim the research article and as a group, summarize the original research study design (i.e., describe the study's sample design, determine if the study was randomized or observational). They were asked to assess the validity of the claims presented in the ``Diner Beware" regarding their research article.

Students   a brief set of slides summarizing their original article using RMarkdown \cite{baumer:2014}. The RMarkdown slides were then shared with the class via RPubs, a platform for web publishing from RStudio (the slides could be submitted in other ways e.g. by emailing the instructor). Lastly, students were given 5-10 minutes to present their findings. The student presentations were intended to improve communication skills as well as allow students to gain experience with technological innovations like RPubs.
	
An example of this process can be presented with the research article titled ``Odors and consumer behavior in a restaurant" \cite{gueguen:2006}. Gu\'eguen and Petr's work analyzed the effect of lemon and lavender scents on the duration of time and the amount of money spent by customers in a restaurant. They carried their study from 8 pm to 11 pm on three Saturdays in May with 88 patrons and hypothesized that lavender is considered a relaxing odor while lemon is a stimulating odor. Another example is presented by the research article titled ``The Impact of Restaurant Table Characteristics on Meal Duration and Spending" \cite{kimes:2004}. In this paper, Sherri E. Kimes and Stephani K. A. Robson assessed how table type and table location can affect average spending per minute (SPM) of a customer. 

Lam summarizes Gu\'eguen and Petr's research article with only the following: ``particular scents also have an effect: diners who got a whiff of lavender stayed longer and spent more than those who smelled lemon, or no scent" \cite{lam3:2015}. Similarly Lam provides a terse summary for Kimes and Robson's article stating that ``Diners at banquettes stayed the longest...Diners at bad tables-next to the kitchen door, say-spent nearly as much as others but soon fled." The students were asked to reconcile these statements with the conclusions presented in the original research articles. 

\section*{Results}

The activity was conducted with Introductory and Intermediate statistics students at Amherst College in Fall 2015 and Spring 2016 academic semesters. The Amherst College Institutional Review Board (IRB) approved this study. On average, 20-25 students in each class engaged in the activity. Approximately 80 minutes were allotted to the activity.

In summary, students correctly identified basic conceptual elements in the original studies' designs. These elements include sample size, the research question, conclusion, and the classification of the study as observational or randomized. Many students were skeptical of the brief claims about the original studies given in the ``Study of Studies."

For example, student work correctly identified Gu\'eguen and Petr's sample size of patrons from a small pizzeria in Brittany, France. The students also describe how ``lavender, but not lemon, increased the length of stay of customers and the amount of purchasing," which indicates that the students' picked up on Gu\'eguen and Petr's hypothesis and research conclusions. 

The students criticized the way the conclusions were portrayed in the ``Diner Beware" article; the students identify how the ``Diner Beware" summary does not account for the possibility of ``cultural bias/
geographical bias." Geographical bias stems from the fact that the study was only conducted in a small town in France and so the conclusions regarding scent and customer spending behavior may not generalize well to people of non-French heritage or individuals from urban areas. 

In their article, Gu\'eguen and Petr acknowledge that a small sample size and the use of only one restaurant are limitations of their study; students picked up on these caveats. Student also recognized how ``limiting the study to three Saturdays in May between 8pm-11pm further creates sampling bias (targets a specific population)." Daytime and weekday visitors are evidently not represented. Moreover since there was no replication, it is highly possible that another factor may have confounded the results. 

Student analysis of Kimes and Robson's article also revealed comprehension of the research's design. In their analysis, the students correctly identify the sample size of 1,413 and the single-blinded nature of the study, since in students' words, ``the participants did not know the true nature of the experiment." The students expressed skepticism regarding the causal statements made in Lam's article regarding Kimes and Robson's study, considering the observational nature of the study and the fact that Kimes and Robson ``excluded some information, like the bar and patio seating" and that they ``only took data from busy times." Kimes and Robson's limitations stem from the fact that they only used one restaurant to draw conclusions, a shortcoming that relates to the limitation students picked up in regards to the limited focus of the study (i.e., inattention paid to less busy hours). Another student group summarized the limitation of Lam's synthesis as the inability to generalize the original research's findings since the ``conclusion for this specific restaurant may not apply to all restaurants." Students' propensity for critique allows for challenging conventions which produces a skepticism and curiosity driven outlook. This outlook though may need to challenged or reconsidered since student skepticism may be excessive.  

\section*{Discussion}

We described an activity that linked summaries of research studies with published 
scientific papers. In general, students accurately reported the original research's study design, in particular, the study's sample size, whether it was observational or experimental, and the general hypothesis as well as the overarching conclusions. Students were often critical of the extremely terse representations of the original research by Lam's ``Diner Beware" article in \emph{The Atlantic}'s ``Study of Studies" column. This is not surprising given that the goal of the ``Study of Studies" is to introduce provocative or idiosyncratic research findings and not to comprehensively review or assess them. 

Time-permitting, the instructor might spend some time debunking misplaced criticism, ensuring that the students have a thorough understanding of the original research, can acknowledge credible published findings, and not develop ``knee-jerk" skepticism. 

Overall, this activity was successfully implemented. It raised awareness about study design as well as secondary representations of original research. The activity can be undertaken with introductory and intermediate statistics students in a single class period and may help improve communication skills by fostering discussion about experimental design. We recommend that the study be undertaken after one or more lectures in study design. Conducting the study after few lectures would provide an informal student assessment and in the process, help reinforce previously-learned study design concepts. 

We believe numerous other articles published in the ``Study of Studies" column could be utilized in the same way as the ``Diner Beware" article. Depending on student interest, other suitable articles include ``Gullible Brains: How our senses influence our thoughts," ``CEOs Behaving Badly: What a chief executive's golf game and handwriting say about his compensation—and his leadership," ``Status Anxiety: What the logos you're wearing really say," and ``The Science of Beer Goggles: Alcohol makes people impulsive, vain, and uncharitable—and it just might help them maintain committed relationships" (see Table \ref{tab:diner} for a comprehensive list of candidate articles). 

\section*{Acknowledgements}

We would like to acknowledge support from the Gregory Call Fund from the Amherst College Dean of the Faculty. We thank Tasheena Narraidoo for her assistance.

\singlespacing
\bibliographystyle{agsm}\bibliography{refsDiner}
\end{document}